\documentclass[twocolumn,a4paper,superscriptaddress,floatfix,showpacs]{revtex4}
\usepackage{amsmath,amssymb,epsfig,color,textcase,hyphenat}

\begin{document}

\title{Unconventional magnetism as a consequence of the charge disproportionation
and the molecular orbital formation in Ba$_4$Ru$_3$O$_{10}$}
\author{S.V.~Streltsov}
\affiliation{Institute of Metal Physics, S.Kovalevskoy St. 18, 620990 Ekaterinburg, Russia}
\affiliation{Ural Federal University, Mira St. 19, 620002 Ekaterinburg, Russia}
\email{streltsov@imp.uran.ru}

\author{D.I.~Khomskii}
\affiliation{II. Physikalisches Institut, Universit$\ddot a$t zu K$\ddot o$ln,
Z$\ddot u$lpicher Stra$\ss$e 77, D-50937 K$\ddot o$ln, Germany}


\date{\today}

\begin{abstract}
The magnetic and electronic properties of Ba$_4$Ru$_3$O$_{10}$ were 
investigated by the ab-initio and model calculations. It is shown that 
nonmagnetic ground state of the one-third Ru$^{4+}$ ions is not due to 
the correlation effects. It is rather caused by the charge 
disproportionation between crystallographically different Ru and
the molecular orbital formation in the Ru's trimer.
\end{abstract}

\maketitle

\section{Introduction \label{intro}}
The complex Ru oxides are widely investigated in the last decades
because of their unusual electronic and magnetic properties. 
The odd-parity pairing was found in the
superconductor Sr$_2$RuO$_4$,~\cite{Nelson2004} while
Ca$_2$RuO$_4$ became one of the model systems for study of the 
orbital-selective Mott transition.~\cite{Gorelov2010}
The integer spins on the Ru$^{4+}$ ions likely form the Haldane 
chains in Tl$_2$Ru$_2$O$_7$, which leads to a drastic decrease of the 
magnetic susceptibility at low temperatures.~\cite{Lee2006}
Similar behavior of susceptibility was recently observed in another
Ru oxide - Ba$_4$Ru$_3$O$_{10}$.~\cite{Klein2011}  This drastic decrease was 
initially attributed to the formation of the unconventional 
antiferromagnetic (AFM) state, but the mechanism of its
stabilization is still unknown.

\begin{figure}[b!]
 \centering
 \includegraphics[clip=false,width=0.5\textwidth]{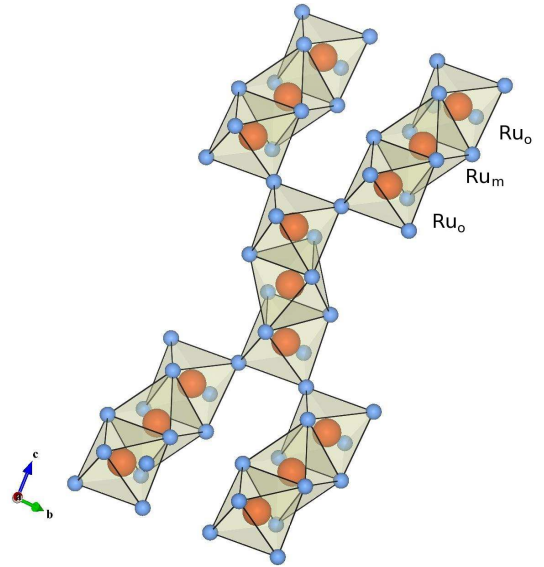}
\caption{\label{Cryst.str}(color online). The crystal structure of 
Ba$_4$Ru$_3$O$_{10}$. Projection on the $bc-$plane. Oxygen ions are
in blue, Ruthenium - in red, Ba - not shown for the simplicity.
Ru$_m$ are in the middle of the trimers, Ru$_o$ -- on the corners.}
\end{figure}

Structurally Ba$_4$Ru$_3$O$_{10}$ is made of the Ru-trimers.
These trimers are formed by three RuO$_6$ octahedra
sharing their faces. Each trimer is connected with four
neighboring trimers (see Fig.~\ref{Cryst.str}) via corners of the 
outer RuO$_6$ octahedra to  build corrugated layers. 
We denote the middle Ru ion in the trimer Ru$_m$, while two outer 
Ru -- Ru$_o$.

In Ba$_4$Ru$_3$O$_{10}$ the Ru ions must have 4+ oxidation state with
4 electrons in the $4d-$shell, since Ba is 2+ and O is 2-. Due to a large $t_{2g}-e_g$ 
crystal-field splitting (see Sec. III for details) these four 
$4d-$electrons should be stabilized in the $t_{2g}$ sub-shell resulting in the 
state with S=1. Indeed, at room temperature 
Ba$_4$Ru$_3$O$_{10}$ was found to be paramagnetic with
the effective moment $\mu_{eff}=2.83$~$\mu_B$ typical
for the S=1 ions,\cite{Klein2011,Dussarrat1996} while the decrease of 
the temperature below  $T_N=105$~K leads to the formation of the 
unconventional AFM state, where one-third of the Ru$^{4+}$ ions
are not ordered (tentatively these are Ru$_m$).~\cite{Klein2011} 

Several models were proposed in Ref.~\onlinecite{Klein2011} to explain
this physical phenomenon. One may think that (1) due to some reason a 
part of the Ru ions stays paramagnetic down to a very low temperature; 
that (2) they are randomly frozen; or that (3) because of a strong 
spin-orbit coupling a nonmagnetic state of the Ru ion with $J=0$ 
is stabilized.

In order to check all these possibilities, the ab-initio band structure 
calculations were performed for this compound. We found that
the true reason for the absence of the magnetic moments on one-third
of the Ru ions is the charge disproportionation 
between two crystallographically nonequivalent Ru 
(Ru$_m$ and Ru$_o$) and formation of the molecular orbital 
in the trimers. In the band-structure picture it is manifested in
the suppression of the nonmagnetic 4d density of states (DOS) of the middle 
Ru$_m$ in the vicinity of the Fermi level.

\section{Calculation details}
For the band structure 
calculations we primary used the pseudo-potential PWscf code.~\cite{Giannozzi2009} 
We utilized generalized gradient 
approximation (GGA) with Perdew-Burke-Ernzerhof version of
the exchange-correlation potential~\cite{Perdew1996} and ultrasoft 
scalar-relativistic pseudo-potentials with nonlinear core correction (for better description
of the magnetic interactions).
The charge density and kinetic energy cut-offs were taken 40 Ry and 
200 Ry, respectively. 664 $k-$points in a full Brillouin zone were 
used in the calculation. 
\begin{figure}[t]
 \centering
 \includegraphics[clip=false,angle=270,width=0.5\textwidth]{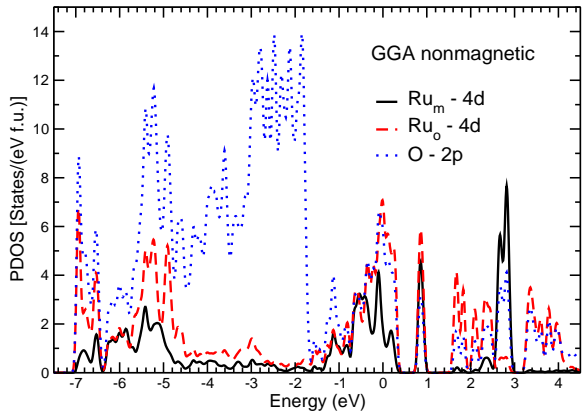}
\caption{\label{GGADOS}(color online). The GGA partial 
density of states, obtained in the PWscf code. Ru$_m$ is the middle
atom in the trimer, while Ru$_o$ are the corner, outer to the trimers, ones.
The Fermi energy corresponds to zero.}
\end{figure}

The generalized gradient approximation with the account of the on-site Coulomb 
repulsion (GGA+U method)~\cite{Anisimov1997} was used to test an importance 
of the electronic correlations. The Hubbard's on-site repulsion parameter $U$ and 
Hund's rule intra-atomic exchange $J_H$ for Ru were calculated in 
Ref.~\onlinecite{Lee2006} to be 3.0~eV and 0.7~eV.

The Wannier function projection was performed within the
linearized muffin-tin orbitals (LMTO) method~\cite{Andersen1984} as it was 
described in Ref.~\onlinecite{Streltsov2005}.

All the calculations were performed for the crystal structure
corresponding to T=10~K.~\cite{Klein2011}

\section{\label{Sec:Res} Nonmagnetic GGA calculations}
We start with the analysis of the results obtained in the
nonmagnetic generalized gradient approximation (GGA), presented 
in Fig.~\ref{GGADOS}. One may see that the oxygen 2p states
are placed primarily from -7 to approximately \nohyphens{-1 eV}. The top
of the valence band is formed by Ru-4d ($t_{2g}$) states.
The Ru-$e_g$ orbitals are above 1.5 eV.
The $t_{2g}-e_g$ crystal field splitting estimated by the 
center of gravity calculation is 3.38 and 3.20 eV for Ru$_m$ and 
Ru$_o$ respectively. Thus one would indeed expect
that such a strong crystal field splitting would prevent
occupation of the $e_g$ orbitals and $d^4$ configuration
of Ru$^{4+}$ would result in the S=1 ground state
of each Ru ion.

However, there is another one quite important feature of 
the nonmagnetic DOS related to
the splitting between the Ru$_m$ and Ru$_o$ 4d states, which
is clearly seen from Fig.~\ref{GGADOS}. The
same center of gravity calculation shows that this 
splitting exceeds 0.13 eV. The Ru$_m$-4d states integrally
turns out to be lower than Ru$_o$-4d due to a larger mean
Ru-O distance. According to Ref.~\onlinecite{Klein2011},
the difference in the mean Ru-O distances for these two
Ru exceeds 0.035 \AA. This is not a small number, since
the difference in ionic radii between Ru$^{4+}$ and 
Ru$^{3+}$ is 0.06 \AA.~\cite{Shannon76} 
As a result one could expect that there would be more d-electrons on 
the middle Ru, i.e. it would have on the average a smaller valence. 
And indeed, according to our nonmagnetic GGA calculations the difference in
charge between Ru$_m$ and Ru$_o$ is 0.52 electron per atom,
which is even larger than the charge disproportionation on 
the Fe sites in the charge ordered phase of 
Fe$_3$O$_4$.~\cite{Leonov04}

\begin{figure}[t!]
 \centering
 \includegraphics[clip=false,angle=270,width=0.5\textwidth]{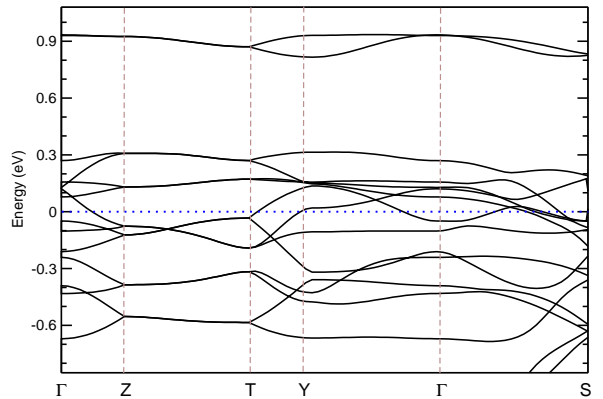}
\caption{\label{GGAbands}(color online). 
The band structure obtained in the nonmagnetic 
GGA calculations (in the PWscf code). The Fermi energy 
corresponds to zero.}
\end{figure}

Thus, Ru$_m$-$t_{2g}$ partial DOS states turns out to 
be almost completely below the Fermi level, 
except a small shoulder and a narrow peak at $\sim$0.9 eV. The detailed
analysis of the band structure, presented in Fig.~\ref{GGAbands},
confirms that the band corresponding to this peak in DOS at ~0.9 eV
is dispersionless. One may
also see that each band in the ZT direction is two times
degenerate, since there are two formula units (f.u.) in the
unit cell. In effect each trimer provides one dispersionless
band at $\sim$0.9 eV.

In order to analyze the formation of this band 
we carried out the calculation within
the local density approximation (LDA) and the LMTO method for which 
one may perform the Wannier function projection procedure.~\cite{Streltsov2005}  The use
of this procedure allows to get the on-site Hamiltonian
matrix for the Ru$-4d$ states and all the hoppings between 
different sites. The projection was performed onto 
thirty Ru$-4d$ orbitals (5 orbitals $\times$ 3 atoms
$\times$ 2 formula units) in the local coordinate system,
where the axis are directed to the nearest oxygens.

The diagonalization of the on-site Hamiltonian in a real
space for each Ru leads to the order of the levels predicted by the 
local distortions. The tetragonal elongation of the Ru$_m$O$_6$ octahedra
results in two low-lying almost degenerate $t_{2g}$-levels
split from the higher $t_{2g}$ orbital by $\sim$330~meV. 
However, a close inspection of the Hamiltonian shows that
the hopping integrals between Ru$_o$ and Ru$_m$ exceed 360~meV, which is,
in contrast to naive expectations, even
larger than the off-diagonal on-site matrix elements
for the $t_{2g}$ orbitals. Thus, the orbitals on which
the Ru$-4d$ electrons tend to localize are determined not only by 
the crystal-field created by the local surroundings (oxygens),
but also by the inter-site matrix elements. This is mainly
due to a face-sharing packing of the RuO$_6$ octahedra in the 
trimers, which allows a direct overlap between $d-$orbitals on the 
neighboring Ru sites. 
Indeed, the nearest neighbors Ru-Ru distance across the common face 
2.55~$\AA$~is even less than that in ruthenium metal, 
2.65~$\AA$.~\cite{Owen37} It is also quite important that
the relevant orbitals here are $4d-$orbitals, which are more extended
than $3d$.~\cite{Goodenough1963}
\begin{figure}[t!]
 \centering
 \includegraphics[clip=false,width=0.3\textwidth]{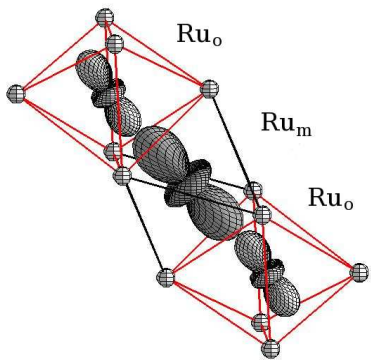}
\caption{\label{a1g}(color online). The molecular orbital which corresponds
to the highest in energy Ru$-t_{2g}$ states (bands at $\sim0.9$~eV
in Fig.~\ref{GGAbands}), obtained by the diagonalization of the 
three-site Hamiltonian with the use of the Wannier function procedure
in the LDA LMTO calculations.}
\end{figure}

In order to take into account the hoppings, we constructed
the large 15$\times$15 Ru$_o$-Ru$_m$-Ru$_o$ Hamiltonian for the 
trimer in a real space.
Diagonalizing this Hamiltonian one obtains that
the lowest and highest in energy are the molecular 
orbitals which have the $a_{1g}$ symmetry with the
largest contribution on the central Ru$_m$ ion. This
is clearly seen in the Fig.~\ref{a1g}, where the
highest in energy  antibonding $a^*_{1g}$ molecular
orbital is plotted. This orbital corresponds to the band, 
which has the energy $\sim0.9$~eV in the ZT direction.

Inspecting Fig.~\ref{a1g}, one may see that the contribution of the 
Ru$_m-4d$ states to the antibonding $a^*_{1g}$ orbital should be the 
largest (the same is actually correct also for the bonding
$a_{1g}$ wave function). This is easy to understand by solving
the three-site ($i=1,2,3$) problem with the single orbital per site in 
the tight-binding approximation. For simplicity the energies of
each orbital are taken the same, $\varepsilon_i = 0$, and the hoppings, 
$t>0$, are nonzero only between the nearest sites. For the linear
cluster of three atoms the Hamiltonian is written as
\begin{equation}
\label{hamiltonian}
H = \left(
\begin{array} {ccc}
0 & t & 0 \\
t & 0 & t \\
0 & t & 0 \\
\end{array}
\right). 
\end{equation}
The energy spectrum consists of the bonding (b), 
antibonding (ab) and nonbonding (nb) orbitals:
\begin{eqnarray}
E_{ab} = t\sqrt 2,&& \psi_{ab} = \frac 12 (\psi_1 -  \sqrt{2} \psi_2 + \psi_3), \\
E_{nb} = 0,&&       \psi_{nb} = \frac 1{\sqrt 2} (\psi_1 - \psi_3), \\
E_{b} = -t\sqrt 2,&&  \psi_{b} =  \frac 12 (\psi_1 +  \sqrt{2} \psi_2 + \psi_3).
\end{eqnarray}
\begin{figure}[t]
 \centering
 \includegraphics[clip=false,angle=270,width=0.5\textwidth]{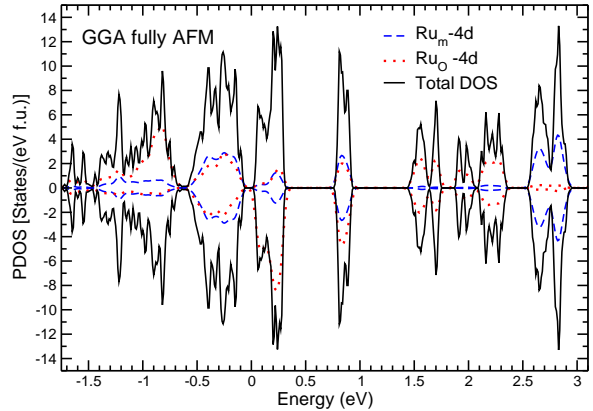}
\caption{\label{GGAmagDOS}(color online). The total and partial density
of states (DOS) in
the GGA calculation for the $\uparrow m \downarrow - \uparrow m \downarrow  $
magnetic configuration (so called fully AFM), when the spins on the Ru$_o$ both 
in the same and in the adjacent trimers are antiparallel.
The results were obtained in the PWscf code. The Fermi energy 
corresponds to zero.}
\end{figure}

As a result one may see that the weight of the states corresponding to the 
middle, $i=2$, atom to the bonding and antibonding wave functions equals to 50\%, 
while they do not contribute to the nonbonding molecular orbital at all.
Thus, the Ru$_m-t_{2g}$ states of the $a_{1g}$ symmetry are pushed away 
from the Fermi level due to the molecular orbital formation. Moreover,
the shift of the rest of the Ru$_m-t_{2g}$ states downwards, with the corresponding
charge disproportionation, leads to the suppression of the Ru$_m$ 4$d$ partial
DOS
in the vicinity of the Fermi level. The Ru$_o-4d$
states in contrast form well defined peak exactly at the Fermi
energy.

The Stoner criteria cannot be applied directly to the partial DOS, but it is 
clear that the gain in the magnetic energy due to the spin splitting 
on Ru$_m$ will be minimal (because of the pseudogap in its partial DOS), 
while the loss of the band energy will be substantial. Thus in the
band picture the 
Ru$_m-4d$ states are not expected to be polarized.
In contrast, the Ru$_o-4d$ states can be easily magnetized. This
is exactly what we observe in the magnetic GGA calculations.

\section{Magnetic GGA calculations}

Several configurations were investigated in our magnetic GGA 
calculations (see Tab.~\ref{TotEnGGA}). The lowest in energy is 
the state when the spins on the Ru$_o$ both in the same and in the adjacent trimers are 
antiparallel ($\uparrow m \downarrow - \uparrow m \downarrow$). The
partial DOS corresponding to this configuration are plotted in Fig.~\ref{GGAmagDOS}.
One may see that there is no splitting between majority
and minority spins of the Ru$_m$. Thus according to our
calculation 1/3 of the Ru atoms indeed are nonmagnetic, exactly
as it was observed in the experiment.~\cite{Klein2011}

\begin{table}[t]
\centering \caption{\label{TotEnGGA}
Total energies, magnetic moments on the Ru$_o$ and Ru$_m$ ions and 
band gaps. Results of the GGA in the PWscf code. Total energies are 
per formula unit (f.u.). The notation $\uparrow m \downarrow$
means that two Ru$_o$ ions are AFM, while moment for the Ru$_m$
can be found in the 3rd column. Symbol ``-'' denotes the bond with
adjacent trimers. The unit cell vectors are chosen in such a way, 
that the type of inter-trimer magnetic order (ferro- or antiferromagnetic) 
for presented configurations is the same for all 4 inter-trimer bonds.}
\vspace{0.2cm}
\begin{tabular}{c|c|c|c}
\hline
\hline
               & Total energy   &   Spin moments     & Band      \\
               & (meV)           & Ru$_o$/Ru$_m$ ($\mu_B$) & gap (eV)  \\
\hline
$\uparrow m \downarrow - \uparrow m \downarrow  $
               & 0    & 0.95/0       & 0.12 \\ 
$\uparrow m \uparrow - \uparrow m \uparrow$
               & 22.7 & 0.98/0.33    & Metal \\ 
$\uparrow m \downarrow - \downarrow m \uparrow$
               & 20.1 & 0.94/0   & Metal\\  
$\uparrow m \uparrow - \downarrow m \downarrow$
               & 61.0 & 0.69/0.15   & Metal\\ 
$0m0 - 0m0$
               & 137.6  & 0/0   & Metal\\ 

\hline
\hline
\end{tabular}
\end{table}

The magnetic moment on the Ru$_m$ appears only
in the configurations with the FM order of the Ru$_o$ in the
trimers. The ferromagnetically ordered spins of the Ru$_o$ create
an exchange field on the Ru$_m$, which slightly magnetize 
these ions. On the band-structure language the appearance of 
some magnetization on Ru$_m$ in a ferromagnetic state can be explained by
the spin splitting of the Ru$_m-4d$ states due to their 
hybridization with the Ru$_o-4d$ states, which as we have shown 
above is large enough.

The magnetic moments on Ru$_o$ are smaller than 2$\mu_B$ expected
for S=1. This is a rather typical situation for the transition metal 
oxides (especially for those based on the 4d and 5d 
elements) and can be attributed to a strong hybridization 
with the oxygens. Because of the hybridization we also
observe non zero magnetic moments on some of the oxygens,
which vary from 0.04 to 0.26~$\mu_B$, depending on the
magnetic configuration and oxygens crystallographic positions.
The largest magnetic moment, 0.26~$\mu_B$, was found on the 
oxygen atoms, which belongs to two neighboring trimers in
the fully FM solution ($\uparrow m \uparrow - \uparrow m \uparrow$).

The analysis of the calculation results presented in Tab.~\ref{TotEnGGA} 
shows that the magnetic contributions
to the total energies of different solutions is not described
by purely Heisenberg terms, like $J_{ij} \vec S_i \vec S_j$. Indeed, if 
one compares the total energies of two solutions with the AFM
order in the trimer (e.g. $\uparrow m \downarrow - \uparrow m \downarrow$
and  $\uparrow m \downarrow - \downarrow m \uparrow$) then the lowest
will be the one with the AFM coupling between trimers. However,
if one analyses another two solutions with the FM order
in the trimer (e.g. $\uparrow m \uparrow - \uparrow m \uparrow$
and  $\uparrow m \uparrow - \downarrow m \downarrow$), it becomes clear that
the lowest will be the one with the FM coupling between trimers.
This is mostly related with the smaller magnetic moment on the Ru$_{m}$
ions in the $\uparrow m \uparrow - \downarrow m \downarrow$ configuration,
and it may be a consequence of molecular orbital formation on trimers".

There are also two other contributions to the magnetic energy in
addition to the conventional Heisenberg model, which have to be taken
into account to describe magnetic properties of Ba$_4$Ru$_3$O$_10$. 
This is the development of the magnetic moments
on Ru$_m$ (when the spins in the trimer are parallel) and on those oxygens, 
which are shared by two trimers (when the spins on the neighbor trimers are
FM ordered). In each case the contribution to the total energy is 
$-IM^2/4$, where $M$ is the magnetic moment on the Ru$_m$ or O ions, 
and $I$ - the Stoner parameter, which is approximately equals to the Hund's 
rule exchange parameter for Ru (0.7 eV) and to 1.6 eV, as it was 
calculated in Ref.~\cite{Mazin1997}

It is important to mention that only fully AFM state,
corresponding to the lowest total energy, is insulating
in the GGA approach, without including electronic correlations
(Hubbard's U). 
This is related to the band narrowing caused by the
antiferromagnetism. The small gap insulating ground state, obtained 
in the magnetic GGA calculations agrees with 
a semiconducting temperature dependence of the electric
resistivity, observed in Ref.~\onlinecite{Klein2011}

\section{Correlation effects}

It is well known that for the description of the electronic
and magnetic properties of the transition metal oxides one
often needs to take into account strong Coulomb correlations,
which can be incorporated in the calculation scheme via the 
LDA+U/GGA+U~\cite{Anisimov1997} or LDA+DMFT~\cite{Anisimov97} 
formalism.

We repeated the total energy calculation for some of the magnetic 
configurations within the GGA+U approximation (as mentioned above,
we took the values U=3~eV and J$_H$=0.7~eV, obtained in 
Ref.~\onlinecite{Lee2006})  and found that the 
account of the Coulomb correlations increases the band gap values,
but does not basically change the order of the total energies for different
magnetic configurations (compare Tab.~\ref{TotEnGGA} and \ref{TotEnGGA+U}). This confirms
that the origin of the unconventional magnetic properties
is not related to the correlation effects, typical for the 
localized electrons, but is largely due to two factors 
discussed above: the formation of molecular orbitals on Ru 
trimers and charge redistribution, with the increase of d-electron 
number on the middle ruthenium, Ru$_m$.

\begin{table}
\centering \caption{\label{TotEnGGA+U}
The total energies of the different magnetic solutions, 
as a result of the GGA+U calculations in the PWscf code. Total energies are
given per formula unit. The notations are the same as in Tab.~\ref{TotEnGGA}.}
\vspace{0.2cm}
\begin{tabular}{c|c|ccccc}
\hline
\hline
               & Total energy  & Band gap                  \\
               &  (meV)        & (eV)                  \\
\hline
$\uparrow m \downarrow - \uparrow m \downarrow  $ & 0     & 0.26 eV\\ 
$\uparrow m \uparrow - \uparrow m \uparrow$       & 18.3  & Metal  \\ 
$\uparrow m \downarrow - \downarrow m \uparrow$   & 19.8  & 0.09 eV\\ 
$\uparrow m \uparrow - \downarrow m \downarrow$   & 86.4  & Metal  \\ 
$0m0-0m0 $                                        & 284.6 & Metal  \\ 
\hline
\hline
\end{tabular}
\end{table}

\section{Summary}
Summarizing, on the basis of ab-initio and model calculations we obtained 
the explanation of the unusual magnetic properties of Ba$_4$Ru$_3$O$_{10}$, 
which consist of the Ru trimers coupled via corner-shared oxygens. The ground state 
is found to be an antiferromagnetic insulator, with antiferromagnetic ordering both 
within and between Ru trimers. The most surprising fact -- the nonmagnetic nature 
of the middle Ru ions in each trimer -- is explained to be a result of a 
combined action of the formation of molecular orbitals in the Ru trimers and of 
charge redistribution, with the increasing $d-$states occupation on the middle 
Ru which is sandwiched between two other Ru ions with antiparallel spins. 
These unusual magnetic properties of Ba$_4$Ru$_3$O$_{10}$ are well explained 
by the band picture, and the electron correlations do not play significant 
role in the formation of this magnetic state.

\section{Acknowledgments}
We thank Igor Mazin and Zlata Pchelkina for the fruitful discussions 
about band magnetism. 
This work is supported by the Russian Foundation for Basic Research 
via  RFFI-10-02-96011 and RFFI-10-02-00140, by the Ural branch of Russian
Academy of Science through the young-scientist program, the Ministry of 
education and science of Russia  (grant 12.740.11.0026), by
the German projects SFB 608, DFG GR 1484/2-1, FOR 1346, and by the 
European network SOPRANO.

\bibliography{../library}

\end{document}